\documentclass[twocolumn,showpacs,pre]{revtex4}
\usepackage{graphicx}
\usepackage{dcolumn}
\usepackage{amsmath}
\begin{document}
\title{A model for gelation with explicit solvent effects: Structure and
dynamics}
\author{Michael Plischke}
\address{Physics Department, Simon Fraser University,
         Burnaby, British Columbia, Canada V5A 1S6}
\author{B\'{e}la Jo\'{o}s}
\address{Ottawa Carleton Institute of Physics, University of Ottawa Campus,
Ottawa, Ontario, Canada K1N-6N5}
\author{D. C. Vernon}
\address{Physics Department, Simon Fraser University,
        Burnaby, British Columbia, Canada V5A 1S6}
\date{\today}
\begin{abstract}
We study a two-component model for gelation consisting of $f$-functional
monomers (the gel) and inert particles (the solvent). After equilibration as
a simple liquid, the gel particles are gradually crosslinked to each other
until
the desired number of crosslinks has been attained. At a critical crosslink
density
the largest gel cluster percolates and an amorphous solid forms. This
percolation process is different from ordinary lattice or continuum percolation
of a single species in the sense that the critical exponents are
new. As the crosslink density $p$ approaches its critical value
$p_c$, the shear viscosity diverges: $\eta(p)\sim (p_c-p)^{-s}$ with
$s$ a nonuniversal concentration-dependent exponent.
\end{abstract}


\maketitle \narrowtext
\section{Introduction}\label{sec:intro}
It is generally accepted that percolation is an essential aspect
of gelation or vulcanization --- it is doubtful that even in a
highly entangled melt of long polymers a nonzero value of the
static shear modulus could exist in the absence of an infinite
connected network. However, percolation has usually been studied
in rather special limits. Site and bond percolation of a single
species on regular lattices are very well characterized and
off-lattice percolation seems to present no new
features \cite{DS94}, at least insofar as critical behavior is
concerned. More closely related to real gels are the so-called
correlated percolation models where the distribution of crosslinks
is drawn from a Boltzmann distribution appropriate for a nearest
neighbor lattice gas \cite{SCA82}. Except at special points in the
phase diagram these models are also in the universality class of
the simple percolation problem. In our previous work on transport
properties near the gel point \cite{MP01}, we have also used a
simple one-species percolation process to produce the incipient
gel. We found that the shear viscosity diverges as the percolation
concentration $p_c$ is approached according to $\eta(p)\sim
(p_c-p)^{-s}$ with $s\approx 0.7$. This value of the exponent $s$
is in excellent agreement with a prediction of de Gennes based on
a superconductor-normal conductor analogy \cite{dg79} and with
recent analytical work on a Rouse model \cite{zipp1}. It is also
reasonably close to some experimental results for $s$ \cite{adam}
but quite different from that produced by another set of
experiments $1.1\leq s\leq 1.3$ \cite{colby}. Thus it seems
reasonable to ask if different versions of the crosslinking
process might produce significantly different cluster size
distributions from percolation and, consequently, different
rheological properties.\par 
Gelation often occurs in the presence
of a solvent and over some period of time rather than
instantaneously, as in the usual percolation models. To simulate
this feature, we have considered a two-species model consisting of
a fraction $c$ of $f$-functional particles that are eligible to
bond irreversibly to others of the same kind. The remaining
particles are inert and function as a background liquid through
which the gel particles and clusters diffuse. Crosslinking occurs
in stages: the equations of motion of all the particles are
integrated forward for a fixed number of time steps between
crosslinking attempts and this process is continued until the
desired number of crosslinks is attained. At a critical
concentration of crosslinks, $p_c$, (in the thermodynamic limit)
the largest cluster percolates and an amorphous solid forms. For
this process one can calculate the usual static or geometrical
quantities used to characterize percolating systems, {\it e.g.},
the fraction of particles on the `infinite cluster',
$P_\infty(p)\sim (p-p_c)^\beta$, the mean mass of finite clusters,
$S(p)\sim |p-p_c|^{-\gamma}$, the fraction of samples percolating
$f(p)$, the cluster size distribution
$n(m,p)=m^{-\tau}\phi(m|p-p_c|^{1/\sigma})$ where $m$ is the mass of a cluster
and the radius of gyration $R_g(m)\sim m^{1/D}$  where $D$ is the fractal
dimension of the clusters. For simple percolation
processes, $\tau\approx 2.18$, $\sigma\approx 0.45$ and these two
exponents determine the others through scaling relations
\cite{DS94}. Here we find, at least for small $c$, that the
cluster size distribution, even at $p_c$, is not well described by
a simple power law. However, the other static quantities listed
above do display power law behavior near $p_c$ and a standard
finite-size scaling analysis provides a very good collapse of our
data. Moreover, the hyperscaling relation $2\beta+\gamma=d\nu$,
where $d=3$ is the dimensionality and $\nu$ the correlation length
exponent, is satisfied. This suggests that this percolation
transition is fundamentally describable in terms of a fixed point
with two (at least) relevant scaling fields. As the percolation
point is approached from below, the shear viscosity diverges
according to $\eta(p)\sim (p_c-p)^{-s}$.  In contrast to our
previous work on a model without solvent, we find values of $s$ in
the range $0.3 \leq s(c) <0.45$ as compared with $s\approx 0.7$.
These results suggest that the critical behavior of transport
coefficients of systems close to the gel point is
nonuniversal.\par The structure of this article is as follows. In
section \ref{sec:model} we describe the present model and
simulation procedures in more detail. The geometric properties of
the system are discussed in section \ref{sec:percolation} and the
data on the shear viscosity are presented in section
\ref{sec:visco}. We conclude with a brief summary and discussion
in section \ref{sec:discu}.

\section{The Model}\label{sec:model}
We consider a system of $N$ particles
in three dimensions, all of which interact with each other through
the soft-sphere potential $V(r_{ij})=\epsilon(\sigma_0/r_{ij})^{36}$ 
for $r_{ij}<1.5\sigma _0$ \cite{powles} with
$\sigma _0=1$ and $k_BT/\epsilon=1$. We simulated systems at a
volume fraction $\Phi=\pi\sigma _0^3N/6V =0.4$ which is well below
the liquid-solid coexistence density. In the absence of any other
interactions, this system would be a simple three-dimensional
liquid. We initially place the particles on a simple cubic lattice
that fills the computational box. We then randomly select $N_{gel}=cN$ particles
to be the gel forming component. After equilibration of the system with Brownian
dynamics, with periodic boundary conditions, for 10000 time steps we begin the
crosslinking process. At this point, the calculation proceeds via {\it
conservative} molecular dynamics (MD) so as to allow hydrodynamic modes to
develop. Here we use a time step $\delta
t=0.005\sqrt{m\sigma_0^2/\epsilon}$.  In the smallest system,  crosslinking is
carried out one bond at a time. A
single gel particle is randomly selected and all other gel particles within a
distance of
$1.2\sigma _0$ are identified. One of the particles in this list is randomly
selected and bonded irreversibly to the central particle through the tethering
potential $V_{nn}(r_{ij})=\frac{1}{2}k(r_{ij}-r_0)^2$ with
$k=5\epsilon/\sigma _0^2$ and $r_0=(\pi/6\Phi)^{1/3}\sigma _0$.
Each gel particle is allowed to bond to no more than six others
and bonding between any pair of particles occurs at most once. The
configuration of the entire system is then updated for 100
time steps and the entire bonding process is repeated
until $3pN_{gel}$ crosslinks have been added. The parameter $p$ is
analogous to the occupation probability in a bond percolation
process on the simple cubic lattice. In larger systems, the number of
crosslinks added in the bonding steps is scaled by the system size in order to
keep the crosslinking rate per gel particle constant.\par 
The parameters in the
potentials and the total volume fraction
$\Phi$ are the same as in our previous work \cite{MP01}. The differences are that
in this earlier work all particles were considered to be gel particles and
that the crosslinking was done instantaneously, at $t=0$, when the
particles were on the vertices of a cubic lattice and thus all
structural properties were those of percolation in three
dimensions. The present model is similar in some ways to a model
discussed by Gimel {\it et al.} \cite{gimel95} and Hasmy and
Jullien \cite{hasmy96} who studied percolation in the context of
diffusion-limited cluster-cluster aggregation using Monte Carlo
methods. Their model differs from ours in that it is a lattice
model, in the details of the crosslinking process, in the lack of
solvent and in the nature of the cluster dynamics. In Monte Carlo
simulations, one is forced to arbitrarily choose the
mass-dependent diffusion constant $D(m)$ whereas in our molecular
dynamics calculations it is determined by the existing structure
and the interparticle forces. In the regime that is of interest here,
{\it i.e.}, high enough gel density that percolation is possible,
these authors find the critical behavior of ordinary
percolation.\par 
In a separate set of runs, we calculate the
stress-stress autocorrelation function and, through the
appropriate Green-Kubo formula, the shear viscosity. Equilibration
and crosslinking are carried out as described above and the
calculation of the viscosity is again done with conservative
MD.\par 
The adjustable parameters in our calculations are the gel
fraction $c$, the crosslink density $p$ and the system size. Here
we report results for $c=0.2$, $0.3$ and $1.0$. Calculations for
other values of $c$ are in progress and will be reported in a
future publication \cite{MP02}. We parametrize the size of our
system in terms of the dimensionless length $L=N^{1/3}$ where $N$
is the total number of gel and solvent particles. Because the
crosslinking process is itself quite time consuming, we are able
only to simulate systems up to size $L=32$ (32768 particles) and
this makes our estimates of critical exponents rather imprecise. A
second factor contributing to the uncertainty in critical
exponents is that we need to determine the critical crosslink
density $p_c$ for each value of $c$ whereas for lattice
percolation this number is known to high accuracy. We next discuss
the static (geometric) properties of our model.\par

\section{Percolation}\label{sec:percolation}
\begin{figure}
\resizebox{225pt}{147pt}{\includegraphics{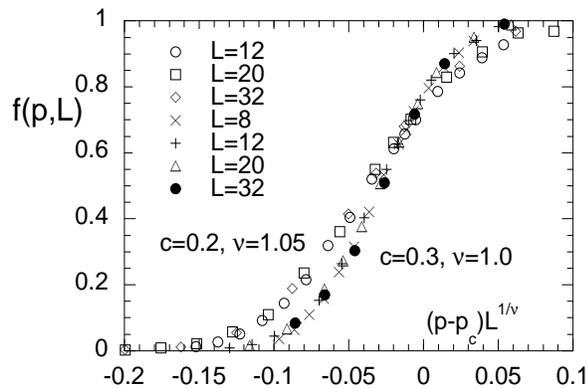}}
\caption{Fraction of samples percolating for $c=0.2$ and $c=0.3$
as function of the scaled crosslink concentration
$x=(p-p_c)L^{1/\nu}$ for $8\leq L\leq 32$. The values of the
exponent $\nu$ used are 1.05 for $c=0.2$ and 1.0 for $c=0.3$. For $L=8$
we have simulated 20000 independent crosslinkings at each $p$; for $L=32$
the data are derived from 3000 samples for each $p$.}
\label{fig1}
\end{figure}

The critical concentration $p_c$ at which percolation occurs in
the thermodynamic limit $L\to\infty$ is accurately estimated from
the intersection of curves $f(p,L)$ as function of $p$ for
different values of $L$. Here $f(p,L)$ is the fraction of samples
percolating in a system of size $L$ at crosslink concentration
$p$. For the two cases of interest here, $c=0.3$ and $c=0.2$, we
find $p_c=0.3165\pm0.0005$ and $p_c=0.3735\pm0.001$. Once $p_c$
has been determined, the correlation length exponent $\nu$ can be
estimated from the collapse of the data for the function $f$ when
plotted as function of $(p-p_c)L^{1/\nu}$. We show this collapse
of the data for $c=0.2$ and $0.3$ in figure (\ref{fig1}). For
$c=0.3$, the best collapse of the data for $8\leq L\leq 32$ is
obtained for $\nu=1.0$ which should be compared to the
three-dimensional percolation result $\nu=0.88$. For $c=0.2$,
finite-size effects are more pronounced and the data for $L=8$
have been excluded. For this case, the best collapse of the data
is obtained for $\nu=1.05$. This method of estimating a critical
exponent is not very accurate but the three-dimensional
percolation value $\nu=0.88$ provides a significantly worse
collapse of the data.\par 
\begin{figure}
\resizebox{225pt}{164pt}{\includegraphics{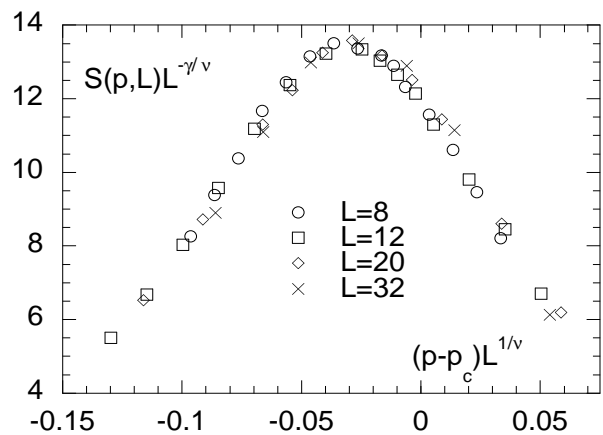}}
\caption{Scaled form of the mean mass of finite clusters
$L^{-\gamma/\nu} S(p,L)$ for $c=0.3$ and $8\leq L \leq 32$. Here
$\gamma/\nu=1.815$ and $\nu=1.0$.}
\label{fig2}
\end{figure}

\begin{figure}
\resizebox{225pt}{164pt}{\includegraphics{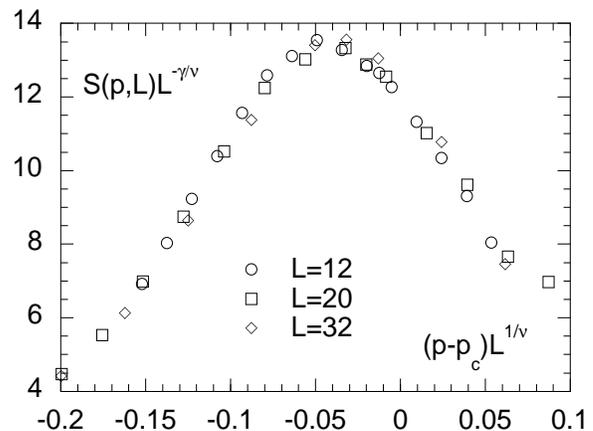}}
\caption{Same as figure (\ref{fig2}) in this case for $c=0.2$ with
$\gamma/\nu=1.80$.} 
\label{fig3}
\end{figure}

We next discuss the mean size of finite
clusters because this data provides an unbiased estimate of the
ratio $\gamma/\nu$. In the thermodynamic limit, $S(p)\sim |p-p_c|^{-\gamma}$ 
with $\gamma\approx 1.8$ for $d=3$ percolation.
For finite $L$, $S(p,L)$ is peaked near $p_c$ with a peak height
that grows as $L^{\gamma/\nu}$. Therefore, rescaling the peak
heights to the same value for different $L$ provides an estimate
of $\gamma/\nu$ that is not affected by errors in either $p_c$ or
$\nu$. Of course, the overall collapse of the data to a universal
curve depends on accurate determination of these two quantities
but the peak height does not. In figures (\ref{fig2}) and
(\ref{fig3}) we show the function $L^{-\gamma/\nu}S(p,L)$ plotted
as function of $x=(p-p_c)L^{1/\nu}$ for the previously determined
values of $p_c$ and $\nu$. The collapse to a universal curve is
quite respectable for both $c=0.3$ and $0.2$ for
$\gamma/\nu=1.815$ and $1.80$, respectively. As above, the data for
$L=8$ have been excluded for $c=0.2$. We note that in the case of
three-dimensional percolation the ratio $\gamma/\nu\approx 2.05$.
Use of this value of $\gamma/\nu$ in figure (\ref{fig2}) would
result in a 40\% difference between the peak heights for $L=32$
and $L=8$.\par 

\begin{figure}
\resizebox{225pt}{164pt}{\includegraphics{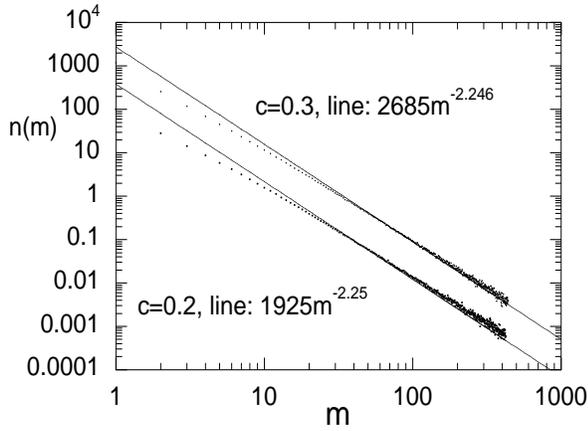}}
\caption{Number of clusters n(m) of mass m for $p\approx p_c$ for
$c=0.2$ and 0.3. The data for $c=0.2$ has been lowered by a factor
of 5 for separation of curves. The straight lines represent fits
to $am^{-\tau}$ with $\tau$ determined by imposing hyperscaling
(see text).} 
\label{fig4}
\end{figure}

\begin{figure}
\resizebox{225pt}{164pt}{\includegraphics{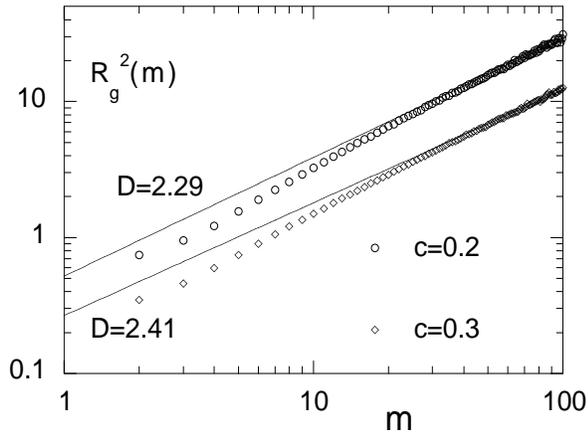}}
\caption{Square of the radius of gyration $R_g^2(m)$ as function
of cluster mass $m$ for $p\approx p_c$ and $c=0.2$ and 0.3.
Straight lines are fits to $R_g^2(m)=am^{2/D}$ with the fractal
dimensions determined by requiring that hyperscaling holds (see
text).} 
\label{fig5}
\end{figure}

\begin{figure}
\resizebox{225pt}{164pt}{\includegraphics{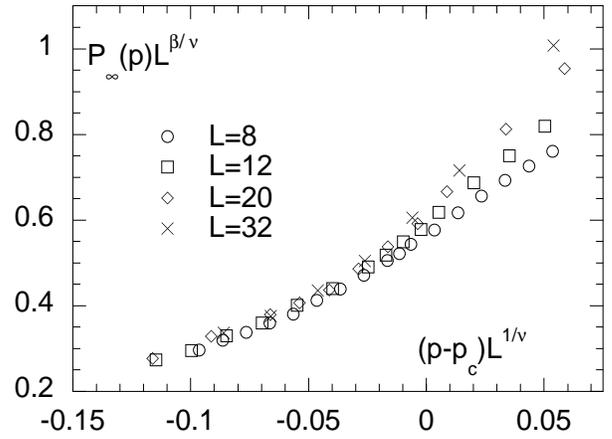}}
\caption{Plot of the scaled form of the order parameter $P(L,p)$
for $p_g=0.3$ and $8\leq L\leq 32$. The exponents are
$\beta/\nu=0.593$ and $\nu=1.0$.}
\label{fig6}
\end{figure}

\begin{figure}
\resizebox{225pt}{164pt}{\includegraphics{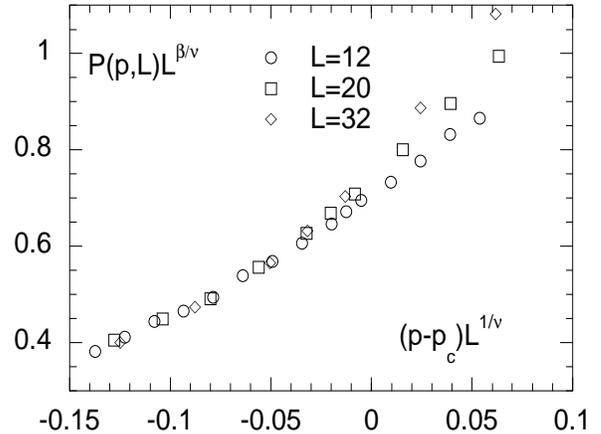}}
\caption{Same as figure (\ref{fig6}) but for $c=0.2$ and
$\beta/\nu=0.6$ and $\nu=1.05$.} \label{fig7}
\end{figure}

In the scaling theory of percolation \cite{DS94},
the ratio $\gamma/\nu=d(3-\tau)/(\tau-1)$, where $d=3$ is the
dimensionality and $\tau$ is the exponent characterising the
cluster size distribution at $p=p_c$. If we enforce this scaling
relation, we obtain $\tau \approx 2.25$ for both $c=0.2$ and
$c=0.3$. Using $\sigma=(\tau -1)/d\nu$, we find
$\sigma(c=0.3)=0.415$ and $\sigma(c=0.2)=0.417$. Using
$D=1/(\sigma\nu)$ for the fractal dimension results in the prediction
$D(c=0.3)=2.41$ and
$D(c=0.2)=2.29$ for the fractal dimensions of the clusters. As
well, the hyperscaling relation $2\beta/\nu=3-\gamma/\nu$ yields
$\beta/\nu=0.593$ and $0.6$ for $c=0.3$ and $0.2$ respectively.
The accuracy of these scaling predictions is tested in figures
(\ref{fig4}) to (\ref{fig7}).\par 
In figure (\ref{fig4}) we show the
number of clusters $n(m)$ of mass $m$ at $p\approx p_c$ for
$c=0.2$ and 0.3 for $L=32$ and $m\leq 400$. For the case $m=1$, we have only
counted the uncrosslinked gel particles. In neither case is the
data well described by a simple power law, in contrast to
percolation on a lattice or in the absence of solvent where the
exponent $\tau\approx 2.18$ is already obtained for $2\leq m\leq 20$. 
A fit to a power law over the range $20\leq m\leq 400$ yields
$\tau=2.13$ for $c=0.2$ and $\tau=2.16$ for $c=0.3$. The straight
lines in figure (\ref{fig4}) are the best fits to the form
$n=Am^{-2.25}$ over the range $20\leq m\leq 400$ and while the fit
is not perfect, the data are not inconsistent with this behavior
in the limit of large $m$.\par 

In figure (\ref{fig5}) we show the
square of the radius of gyration $R_g^2(m)$  as function of $m$
for a system of size $L=32$ together with curves $m^{2/D(c)}$
with $D(c=0.2)=2.29$ and $D(c=0.3)=2.41$ as determined above. The
data again show considerable curvature but the fit to the assumed
functional form is reasonable over the range $20\leq m\leq 100$.\par 
Finally, in figures (\ref{fig6}) and (\ref{fig7}) we
display the scaled form of $P(L,p)$, the probability that a gel
particle is part of the percolating cluster using the predicted
exponent ratios $\beta/\nu=0.593$ for $c=0.3$ and $\beta/\nu =0.6$
for $c=0.2$. These two figures present the least impressive
collapse of data to a universal curve, especially at the larger
values of $P$. One can improve the collapse by different choice of
$\beta/\nu$ and $\nu$ but at the expense of violating
hyperscaling. We also note that the data for the two largest
values of $L$ are reasonably close to each other over the entire
range of $x$.\par 

We have also carried out a limited number of simulations for $c=0.5$
and $c=1.0$ with the crosslinking process described above. In both
cases, the critical exponents and the cluster size distributions
are entirely consistent with ordinary three-dimensional
percolation. This suggests that either there is a critical gel
fraction \cite{hasmy2} below which the geometric properties of the
clusters are described by continuously varying exponents or that
the apparent variation of the exponents with $c$ described above
is a finite-size artefact. Only simulations of larger systems can
resolve this issue.\par

\section{Shear Viscosity}\label{sec:visco}
\begin{figure}
\resizebox{225pt}{164pt}{\includegraphics{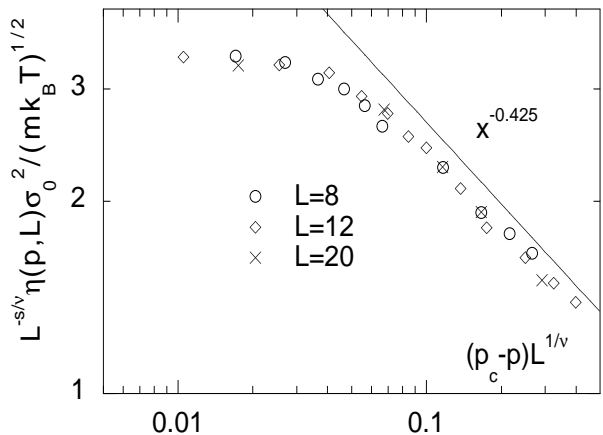}}
\caption{The dimensionless shear viscosity $\sigma_0^2\eta(L,p)/(mk_BT)^{1/2}$
for $c=0.3$ times $L^{-s/\nu}$ plotted as function of $x=(p-p_c)L^{1/\nu}$ with
$s/\nu=0.425$ and $\nu=1.0$. The straight line represents the function
$x^{-s/\nu}$ (see text).} 
\label{fig8}
\end{figure}

We have calculated the shear viscosity for systems up to size
$L=20$ as function of the crosslink density $p$ for $c=0.3$ and
for $L=12$ for $c=0.2$. Systems are equilibrated as a liquid,
crosslinked as described above and then evolved by constant energy
MD for 40000 or 80000 time steps, depending on the crosslink
density. Here we have typically used 500 to 2000 different realizations of the
crosslinks at each $p$. We calculate, as in
\cite{MP01}, the stress-stress autocorrelation function $C_{\sigma\sigma(t)}={1
\over
3}\sum_{\alpha<\beta}\langle\sigma_{\alpha\beta}(t)\sigma_{\alpha\beta}(0)\rangle$
where
\[\sigma_{\alpha\beta}=\sum_{i=1}^Nmv_{i\alpha}v_{i\beta}-\sum_{i<j}
\frac{r_{ij\alpha}r_{ij\beta}}{r_{ij}}V'(r_{ij}) \]
are elements of the stress tensor. Here the sum is over both gel and
solvent particles and $V'$ is the derivative of the pair potential
between particles $i$ and $j$.  The analysis of the stress-stress
correlation function has been described in \cite{MP01} and is done in the same
way here. As $p\to p_c$, $C_{\sigma\sigma}$ decays extremely slowly and is
fitted, at long times, to a stretched exponential. The static shear
viscosity is then obtained from the appropriate Green-Kubo formula \cite{hansen},
\[\eta=\lim_{t_{max}\to\infty}\frac{1}{Vk_BT}\int_0^{t_{max}}
C_{\sigma\sigma}(t)\,.\] 
The results, for $c=0.3$ are shown in finite-size scaled form
in Fig. (\ref{fig8}) where $L^{-s/\nu}\eta(L,p)$ is plotted as
function of the scaled concentration $x$ \cite{note}. In contrast
to our previous result for $c=1$ and instantaneous crosslinking
where we found $s\approx 0.7$, we find that $s\approx 0.425$
provides an excellent collapse of the data with $\nu=1.0$. We note
that, outside the critical region, consistency of the finite-size
scaling ansatz requires the scaled viscosity to vary as
$x^{-s/\nu}=x^{-0.425}$ and it is clear that the data are
consistent with this behavior.\par 
\begin{figure}
\resizebox{225pt}{150pt}{\includegraphics{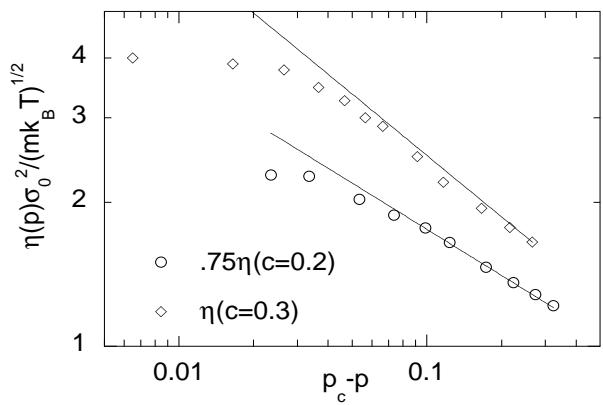}}
\caption{The dimensionless shear viscosity $\sigma_0^2\eta(p)/(mk_BT)^{1/2}$ for
$L=12$ and $c=0.2$ and 0.3 plotted as function of $(p_c-p)$. The straight line
represents the function $(p_c-p)^{-s}$ with $s=0.425$ for $c=0.3$ and $s=0.3$
for $c=0.2$.} 
\label{fig9}
\end{figure}
We have also calculated the shear viscosity for $c=0.2$ for $L=12$. The raw data
are displayed in figure (\ref{fig9}) as function of $p_c-p$ together with the
corresponding results for $c=0.3$. Fitting to a power law outside
the critical region produces an exponent $s\approx 0.3$
suggesting, as in the case of the static properties, a variation
of critical exponents with $c$ and an absence of universality.

\section{Discussion}\label{sec:discu}
In this article we have proposed and investigated a new model for
gelation which incorporates a solvent on a microscopic level. For
relatively small concentrations of gel, the geometric properties
of the system close to the gel point seem to depend continuously
on this gel fraction and are, at least for the system sizes
investigated, markedly different from three-dimensional
percolation. In particular, the fractal dimension of the clusters
seems to be smaller than those of percolation clusters and this
more spidery morphology may be responsible for the slower
divergence of the shear viscosity as the gel point is approached.
The change in the exponents controlling the geometric properties
is rather small and further study of larger systems is certainly
necessary to confirm this result. However, the exponent $s$ that
characterizes the divergence of the shear viscosity at the gel
point is reduced by almost a factor of 2 from its value in the
absence of solvent and it is unlikely that this can be attributed
to finite-size effects. In light of this result, it seems
implausible that a single universality class describes the
behavior of transport coefficients and, presumably, the moduli of
the amorphous phase  near the gel point. The considerable
dispersion found in experimental values of the critical exponents
\cite{adam96} is another indicator that this may be the case.\par
In future work we intend to explore this new model in greater
detail. It will be interesting to investigate if the exponent $s$
and the static exponents are tunable by varying the concentration
of the solvent and the solubility of the solute. We also intend to
study diffusion constants as function of cluster size and to
investigate the existence of long time tails. Finally, one of the
original motivations for this model is the existence of a body of
experimental work that has yielded values in the range 1.1 --- 1.3
for the viscosity exponent $s$. Clearly, we have moved further
from this range of values compared to our previous results. If the
cluster size distribution and cluster geometry is the determining
factor in the critical behavior of the transport coefficients then
this indicates that models that produce more compact rather than
more tenuous clusters than those arising from percolation may be
appropriate.

One of us (BJ) thanks the Physics Department at Simon Fraser
University for its hospitality during a sabbatical visit.
We thank Ralph Colby, Paul Goldbart and Sune Jespersen for helpful
discussions. This research is supported by the NSERC of Canada.

\newpage


\begin{references}
\bibitem{DS94} See for example D. Stauffer and A. Aharony, {\it Introduction to
Percolation Theory}, 2nd Edition, (Taylor and Francis, London, 1994).
\bibitem{SCA82} See D. Stauffer, A. Coniglio and M.Adam, Adv. Pol. Sci.
{\bf 44}, 103 (1982) for a review.
\bibitem{MP01} D. Vernon, M. Plischke and B. Jo\'{o}s, Phys. Rev E {\bf64}, 03105
(2001).
\bibitem{dg79} P.G. de Gennes, J. Phys (Paris), {\bf 40}, L197 (1979).
\bibitem{zipp1} K. Broderix, H. L\"{o}we, P. M\"{u}ller and A. Zippelius,
Europhys. Lett., {\bf 48}, 421 (1999); Phys. Rev. E {\bf 63}, 011510 (2001).
\bibitem{adam}M. Adam, M. Delsanti, D. Durand, G. Hild and J.P. Munch, Pure
Appl. Chem., {\bf 53}, 1489 (1981); M. Adam, M. Delsanti and D. Durand,
Macromolecules, {\bf 18}, 2285 (1985); D. Durand, M. Delsanti and J.M. Luck,
Europhys. Lett., {\bf 3}, 297 (1987).
\bibitem{colby} C.P. Lusignan, T.H. Mourey, J.C. Wilson and R.H. Colby,
Phys. Rev. E {\bf 52}, 6271 (1995); J.E. Martin and J. Wilcoxon, Phys. Rev.
Lett., {\bf 61}, 373 (1988); D. Adolf and J.E. Martin, Macromolecules, {\bf 23},
3700 (1990); J.E. Martin, J. Wilcoxon and J. Odinek, Phys. Rev. A {\bf 43}, 858
(1991); J.E. Martin, D. Adolf and J. Wilcoxon, Phys. Rev. Lett., {\bf 61}, 2620
(1988).
\bibitem{powles} J.P. Powles and D.M. Heyes, Mol. Phys. {\bf 98}, 917
(2000). These authors have studied the properties of systems with
a pair potential of the form $V(r_{ij})=\epsilon(\sigma/r_{ij})^n$, for $12\leq
n\leq 288$, including our case $n=36$.
\bibitem{gimel95} J.C. Gimel, D. Durand and T. Nicolai, Phys. Rev. B {\bf
51}, 11348 (1995).
\bibitem{hasmy96} A. Hasmy and R. Jullien, Phys. Rev. E {\bf 53}, 1789 (1996).
\bibitem{MP02} M. Plischke, S. Jespersen and B. Jo\'{o}s, unpublished.
\bibitem{hasmy2} A critical volume fraction appears in the model of
Gimel {\it et al.} \cite{gimel95}, albeit in a rather trivial way:
When the volume fraction is increased above 0.31, the system percolates
instantaneously (in the thermodynamic limit) and the features associated
with cluster-cluster aggregation disappear.
\bibitem{hansen} M.P. Allen and D.J. Tildesley, {\it Computer Simulation of
Liquids} (Oxford University Press, New York, 1987), Chap. 2; J.P. Hansen and 
I.R. MacDonald, {\it Theory of Simple Liquids}, 2nd ed. (Academic Press, New
York, 1986).
\bibitem{note} A fit of the raw data for $L=12$ and $p\leq 0.26$ to the form
$\eta(p)=a(p_c-p)^{-s}$  produces the estimates $p_c=0.310$ and
$s=0.37$, in good agreement with the percolation theory
estimate of $p_c$ and with the finite-size scaling analysis reported above.
\bibitem{adam96} For a review, see M. Adam and D. Lairez in {\it The Physical
Properties of Polymeric Gels}, J.P. Cohen Addad, ed. (John Wiley and Sons,
Ltd., New York, 1996) p. 87.
\end{references}
\end{document}